\documentclass{webofc}
\usepackage[varg]{txfonts}   
\begin{document}
\title{Spin polarization and alignment in heavy-ion collisions}
%
%

\author{\firstname{Subhash} \lastname{Singha}\inst{1}\thanks{\email{subhash@impcas.ac.cn}} 
}

\institute{Institute of Modern Physics Chinese Academy of Sciences, Lanzhou
}


\abstract{%
  We report the measurements of global spin polarization of
  hyperons ($P_{\Lambda}$) from FAIR, RHIC and LHC energies. The $P_{\Lambda}$ continue to follow
  an increasing trend at $\sqrt{s_{\mathrm NN}}$ < 7.7 GeV
  indicating that enormous vorticity prevail even in hadronic
  dominant region. New measurements of local spin polarization
  ($P_{z,2}$) from isobar collisions follow the same pattern observed in previous
  RHIC and LHC energies. An observable motivated by predicted baryonic Spin Hall
  Effect, the net local polarization ($P_{z,2}^{\mathrm{net}}$), shows
  negative value in Au+Au collisions at $\sqrt{s_{\mathrm NN}}$ = 19.6
  and 27 GeV. The first $P_{z}$ measurement with respect to third
  order harmonic in 200 GeV RHIC isobar collisions show a non-zero sine modulation,
  indicating the presence of complex vortical structure in the medium. Surprising new patterns of spin
  alignment ($\rho_{00}$) for various vector mesons ($K^{*0}$,
  $K^{*\pm}$, $\phi$ and $J/\psi$) emerged at both RHIC and LHC
  energies. The large deviation in $\rho_{00}$ of  several species
  poses challenge to the theory.

}
\maketitle
\section{Introduction}
\label{intro}

In non-zero impact parameter heavy-ion collisions, a very large orbital angular momentum
(OAM $\sim 10^{5} - 10^{7} \hbar$) is expected in the centroid of
the participant matter~\cite{becattini}. The part of OAM transferred
to the Quark Gluon Plasma (QGP) can
polarize quarks (anti-quarks) due to ``spin-orbit''
coupling~\cite{liang0}. Further, the polarization of quarks are translated into the
polarization of produced hadrons with non-zero spin along the OAM via
hadronization processes. In the initial stage, we also expect a large magnetic
field ($B$ $\sim 10^{18}$ Gauss)~\cite{kharzeev}, which can induce different spin polarization for quarks 
(anti-quarks) with different magnetic moments. Thus, through the polarization
measurements we can search for the signatures of initial OAM interactions
and $B$-field. These measurements help understand the response of
QGP medium under these extreme initial conditions and also provide a unique opportunity
to probe the spin dynamics of QGP.  

\section{Spin polarization of hyperons}
\label{sec-1}
\begin{figure}[h]
\centering
\includegraphics[width=6.5cm]{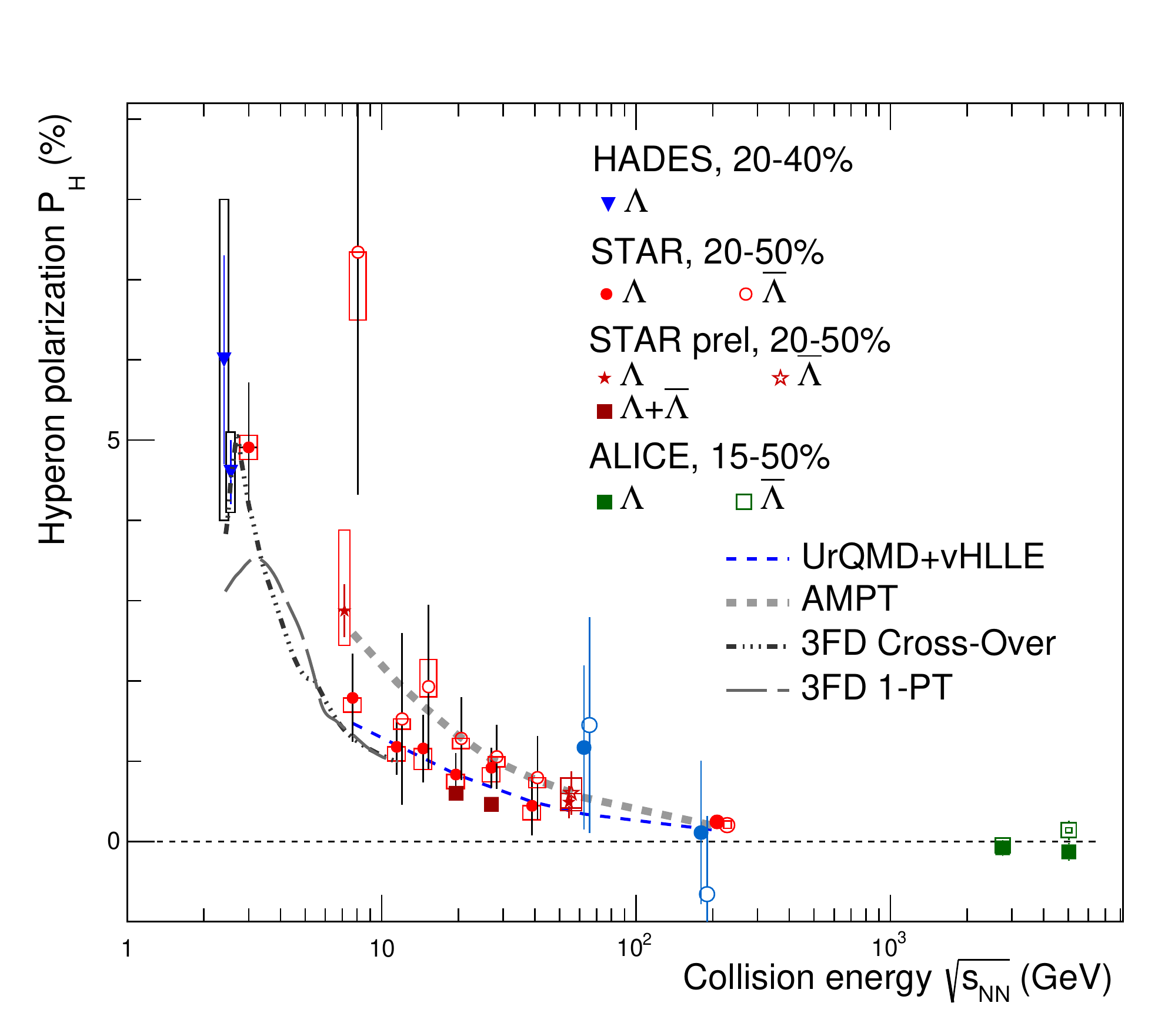}
\caption{Beam energy dependence of global spin polarization of
  $\Lambda$ and $\bar{\Lambda}$
  hyperons~\cite{star_lambda_prc,star_lambda_nature_bes,star_lambda_prc_200GeV,
    alice_lambda_prc,star_lambda_prc_3GeV,hades_lambda_prc}. Experimental
measurements are compared to several models.}
\label{fig-1}    
\end{figure}

Using the parity violating weak decay of hyperons, the global
polarization ($P_{\mathrm{H}}$) can be measured from the angular distribution of the
daughter baryon in parent hyperon's rest frame~\cite{schiling}. In 2007, the STAR experiment set an upper limit on
$P_{\Lambda}$ (< 2\%) in 200 GeV Au+Au collisions~\cite{star_lambda_prc}. A breakthrough happened when STAR reported
the first evidence of positive $\Lambda$ and $\bar{\Lambda}$ spin
polarization in heavy ion collisions from the RHIC Beam
Energy Scan (BES) program. Under the assumption of local
thermal equilibrium and based on hydrodynamic model, the polarization
data averaged over all the BES energies resulted in a vorticity of about
(9 $\pm$ 1) $\times$ $10^{21}$ $/s$. This suggests the most
vortical fluid ever created in a laboratory experiment~\cite{star_lambda_nature_bes}. In the BES
measurements, $P_{\bar{\Lambda}}$ is consistently larger than
$P_{\Lambda}$ across the beam energies. Due to opposite
sign of their magnetic moment, such a splitting in
polarization is expected from the $B$-field. Nevertheless, the precision is not sufficient for a
clear conclusion. Subsequently, a precise non-zero $P_{\Lambda}$ along
with differential measurements of $P_{\Lambda}$ as a function of collision centrality, transverse
momentum and rapidity are reported at the top-RHIC
energy~\cite{star_lambda_prc_200GeV}. The increase of $P_{\Lambda}$
from central to peripheral collisions is consistent with expectation from the
vorticity of the medium. Later, the ALICE experiment at LHC reported $P_{\Lambda}$ consistent with zero
within uncertainties~\cite{alice_lambda_prc}. Recently, the $P_{\Lambda}$ measurements are
extended  in the lower beam energy by the fixed target set up at STAR~\cite{star_lambda_prc_3GeV}
and HADES~\cite{hades_lambda_prc}. A significant $P_{\Lambda}$ of about 5\% is observed at
these energies. The $P_{\Lambda}$ follows a
monotonically increasing trend from 5.02 TeV down to 2.4 GeV as shown
in Fig.~\ref{fig-1}. Various transport and hydrodynamic models can successfully capture this energy dependence
trend. The monotonic increase of $P_{\Lambda}$ is understood as an interplay of
shear flow and baryon stopping. There are scenarios, e.g, migration of
polarization to forward rapidity and lifetime of the QGP etc can also
contribute to $P_{\Lambda}$. Recent model calculations suggested that
$P_{\Lambda} \sim 0$ at $\sqrt{s_{NN}}$ = 2$m_{N}$. The monotonic rise
of $P_{\Lambda}$ at $\sqrt{s_{NN}}$ < 7.7 GeV and a clear increase of $P_{\Lambda}$ from
central to peripheral collisions, in a hadronic dominant
matter, is indeed surprising. It is an open question to the community,
how the hadron gas can support such enormous vorticity? More experimental
measurements and inputs from theory are needed for a better understanding.

\subsection{Global spin polarization in isobar collisions}
\label{sub-sec-1}

\begin{figure}[h]
\centering
\includegraphics[width=6cm]{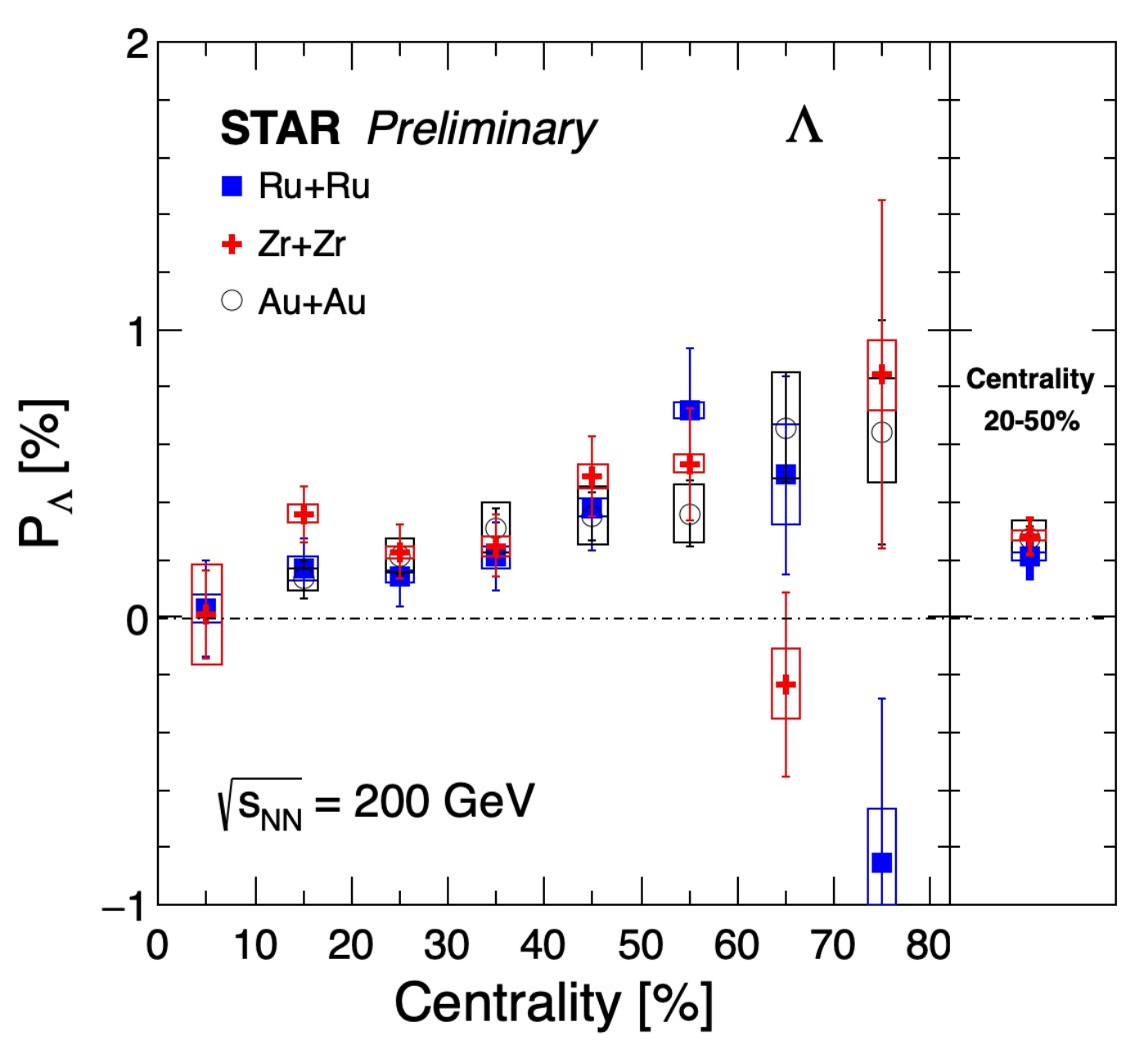}
\includegraphics[width=6cm]{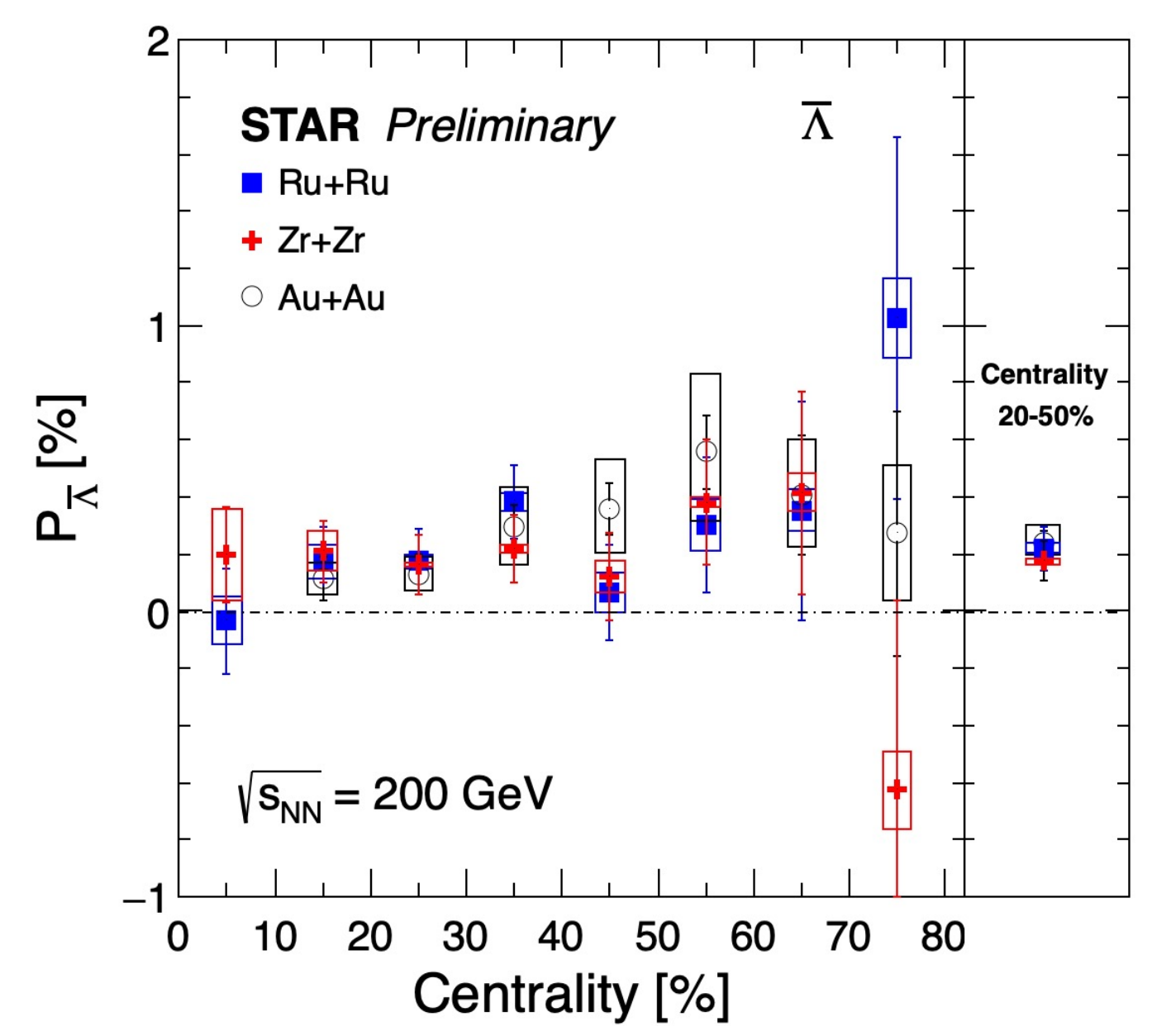}
\caption{Centrality dependence of global spin polarization of
  $\Lambda$ (left) and $\bar{\Lambda}$ (right) in isobar (Ru+Ru and Zr+Zr)
  collisions. Results are compared with the same from Au+Au collisions~\cite{xgou}.}
\label{fig-2}       
\end{figure}

The recent high statistics 200 GeV isobar collisions
($_{44}^{96}$Ru+$_{44}^{96}$Ru and $_{40}^{96}$Zr+$_{40}^{96}$Zr)~\cite{blind_isobar} at
RHIC offer new opportunities for polarization measurements. Due to
larger proton numbers, the Ru+Ru is expected to have larger
initial $B$-field than the Zr collisions. Thus the measurement $P_{\Lambda}$ and
$P_{\bar{\Lambda}}$ polarization in isobar collisions can probe the
initial $B$-field. Furthermore, several model calculations predicted a
collision system size dependence in $P_{\Lambda}$
($P_{\Lambda}^{\mathrm{O+O}} > P_{\Lambda}^{\mathrm{Ru+Ru}} >
P_{\Lambda}^{\mathrm{Au+Au}}$)~\cite{system_size_pol}. The isobar data
can be used to test such dependencies. The STAR collaboration
reported  $P_{\Lambda}$ and $P_{\bar{\Lambda}}$ from isobar
collisions~\cite{xgou}. As shown in Fig.~\ref{fig-2}, the
$P_{\Lambda}$ ($P_{\bar{\Lambda}}$) in isobar collisions followed same
centrality dependence observed in previous Au+Au collisions. No
polarization difference is observed between $\Lambda$ and
$\bar{\Lambda}$ or between the isobar collisions species within the
current precision. However, the results from RHIC BES-II are expected
to provide a precision to probe the $B$-field. No obvious system size
dependence in $P_{\Lambda}$ ($P_{\bar{\Lambda}}$) is observed when
isobar results are compared to large system Au+Au collisions at a same
collision centrality. Results from a smaller system collision (O+O)
data can provide more insights into the system size
dependence of $P_{\Lambda}$. The STAR reported a first time non-zero
polarization for $\Xi$ and $\Omega$ hyperons in 200 GeV Au+Au collisions, which
reinforces the global nature of spin polarization in heavy-ion
collisions~\cite{xi_omg_pol}. More precise measurements of $\Xi$ and
$\Omega$ polarization in future RHIC runs can help testing the species
dependent splitting of polarization due to different spin and magnetic
moments.

\subsection{Local spin polarization of hyperons in isobar collisions}
\label{sub-sec-2}
Recently it is understood that the collective flow in
conjunction with the vorticity of the medium can induce a novel
longitudinal polarization ($P_{z}$) along the beam direction~\cite{Becattini_pz}. The
first non-zero sine modulation of $P_{z}$ for $\Lambda$ ($\bar{\Lambda}$) hyperons with respect to second order flow
harmonics ($\Psi_{2}$) is observed in 200 GeV Au+Au collisions by
STAR~\cite{star_lambda_pz}. Although many hydrodynamic and transport models do excellent job in
capturing the beam energy dependence of the global hyperon
polarization, they failed to explain the pattern in $P_{z}$ with
proper sign, which is termed as a ``sign-puzzle''. Later the same
pattern in $P_{z}$ is also seen in 5.02 TeV Pb+Pb collisions at LHC by ALICE~\cite{alice_lambda_pz}. 
Recently there are many theoretical developments in addressing this puzzle. It is
understood that the addition of a ``shear term'' can explain the
correct sign of $P_{z}$~\cite{shear_pz_sign}. 

\begin{figure}[h]
\centering
\includegraphics[width=6cm]{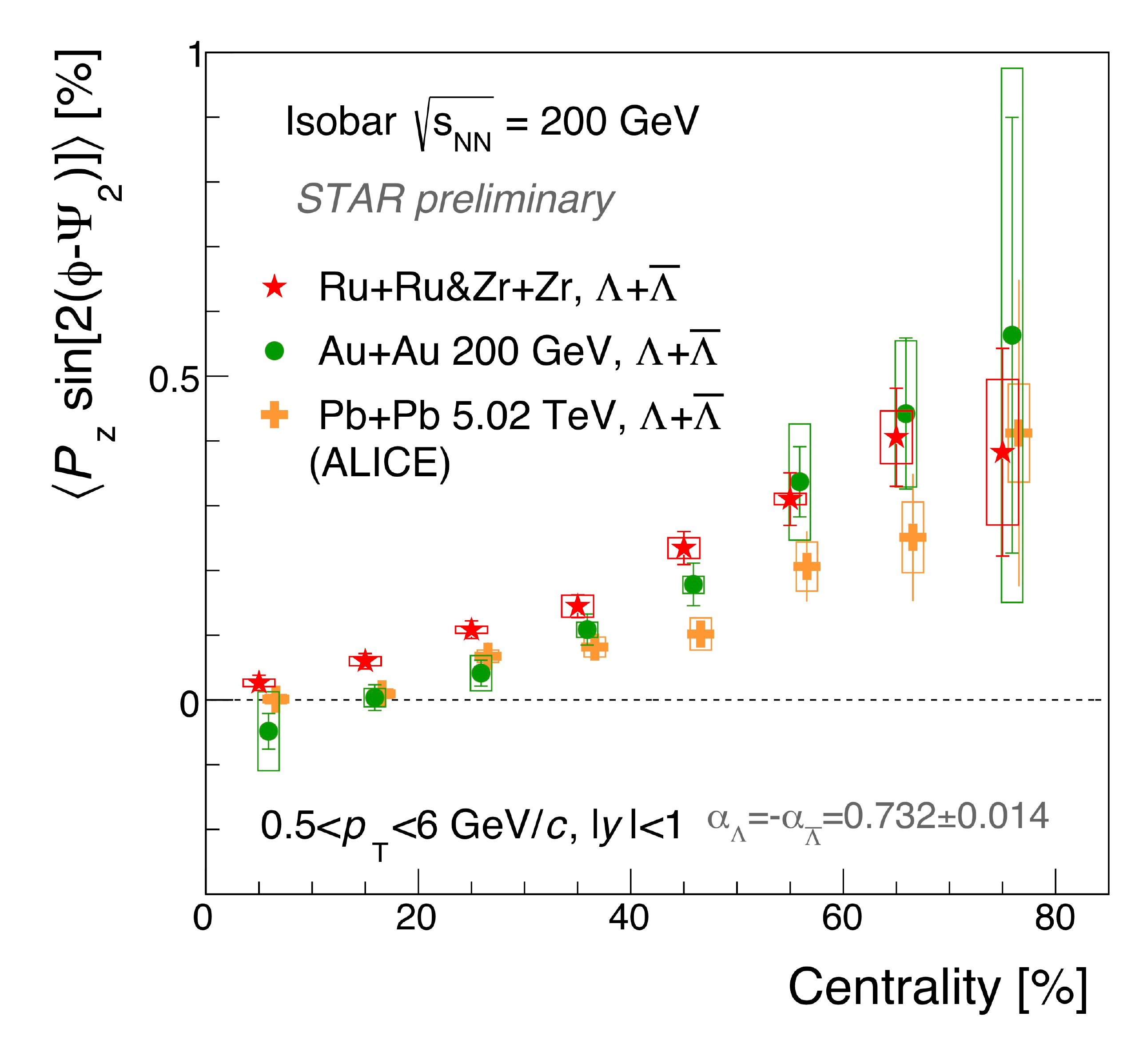}
\includegraphics[width=6.7cm]{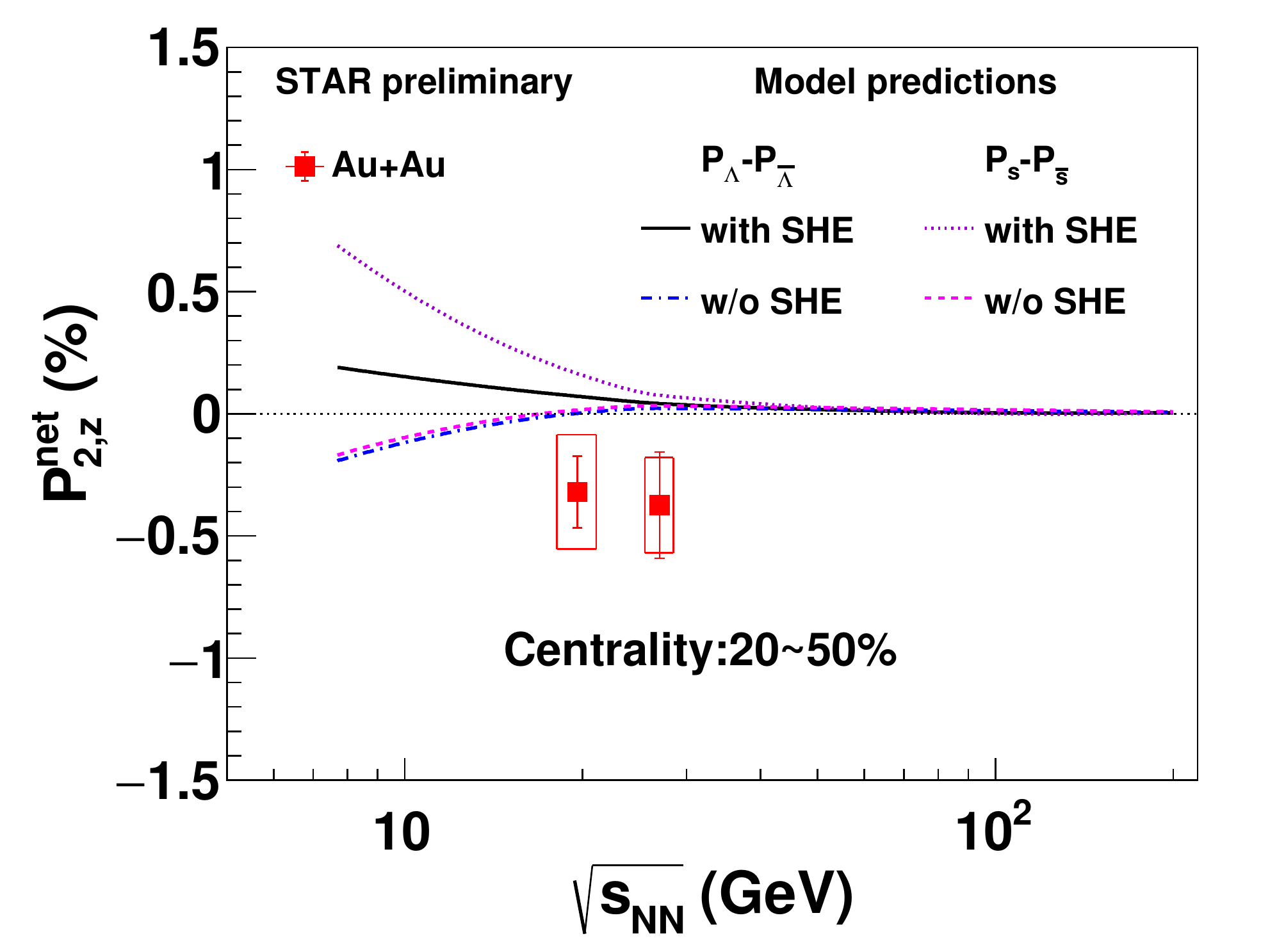}
\caption{Left: Centrality dependence of local spin polarization ($P_{z,2}$)
  of $\Lambda$ hyperons in Ru+Ru, Zr+Zr, Au+Au and Pb+Pb
  collisions. Right: Beam energy dependence of net local polarization
  ($P_{z,2}^{\mathrm{net}}$)  of $\Lambda$  and
  $\bar{\Lambda}$. Results are compared with model predictions.}
\label{fig-3}       
\end{figure}

STAR reported $P_{z}$ with respect to second order harmonics ($P_{z,2}$) from the
isobar collision~\cite{xgou}. The left panel in Fig.~\ref{fig-3} presents the
comparison of isobar results with the same from
Au+Au collisions at 200 GeV~\cite{star_lambda_pz} and Pb+Pb collisions at 5.02 TeV~\cite{alice_lambda_pz}.
There are some hints of system size dependence but no
obvious energy dependence in $P_{z,2}$ is observed. These measurements
in isobar collisions are also extended with respect to third order harmonics
($P_{z,3}$) for the first time and a significant non-zero sine
modulation in polarization is observed~\cite{xgou}. Both $P_{z,2}$ and $P_{z,3}$
show a similar centrality dependence trend, however $P_{z,3}$ is
systematically smaller than $P_{z,2}$ in peripheral collisions.
These local polarization measurements can provide informations
on the complex vortical structures, constrain on initial conditions
and transport parameters in many models.

Recently, it is predicted that the gradient of baryonic chemical 
potential ($\nabla \mu_{B}$) can cause a difference in $P_{z}$ for
$\Lambda$ and $\bar{\Lambda}$ ($P_{z,2}^{\mathrm{net}}$). Having similarity with a spin phenomena
in condensed matter physics, it is termed as baryonic Spin Hall Effect
(SHE)~\cite{she_prediction}. It is predicted that the effect of SHE could be more prominent at lower
beam energy (larger baryonic chemical potential) and it could produce a positive monotonically increasing 
$P_{z,2}^{\mathrm{net}}$ with decrease in beam energy. On the
contrary, absence of SHE is predicted to generate a similar monotonic pattern
but with a negative sign in $P_{z,2}^{\mathrm{net}}$. Abundant
production of $\Lambda$ ($\bar{\Lambda}$) and a large baryonic
chemical potential provide an ideal environment to look for
such novel effects at the RHIC BES range. The STAR experiment reported the net local
polarization $P_{z,2}^{\mathrm{net}}$ in mid-central Au+Au collisions at 19.6 and
27 GeV as shown in right panel of Fig.~\ref{fig-3}~\cite{huqiang}. The sign of $P_{z,2}^{\mathrm{net}}$ is negative
and consistent with absence of SHE. The $P_{z}$ measurements from lower
beam energies ($\sqrt{s_{\mathrm NN}} < $ 19.6 GeV) will be useful in testifying the presence of SHE.

\section{Spin alignment of vector mesons}
\label{sec-2}
\begin{figure}[h]
\centering
\includegraphics[width=6.5cm]{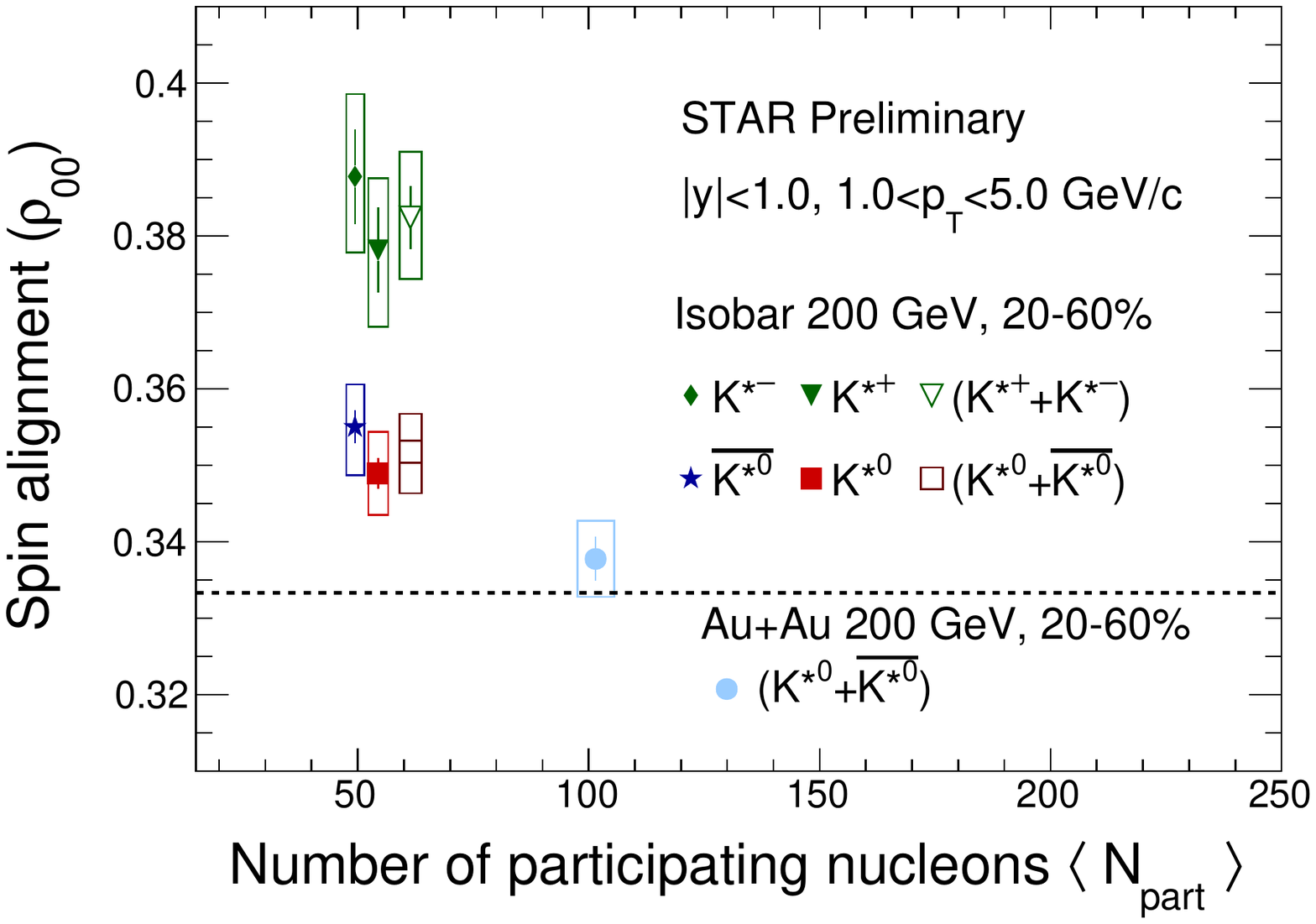}
\includegraphics[width=6.2cm]{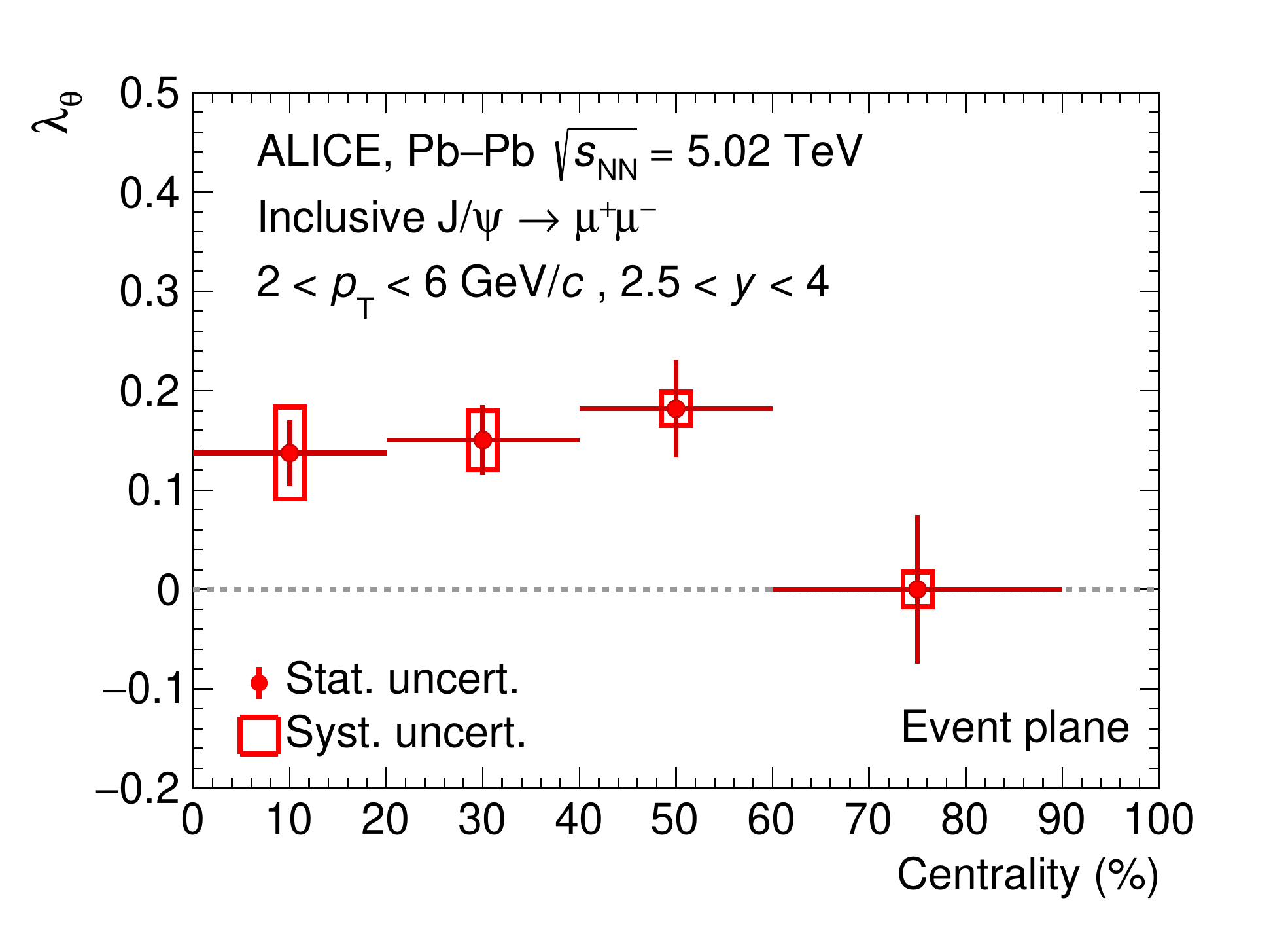}
\caption{Left: Spin alignment ($\rho_{00}$) of $K^{*0}$ ($\overline{K}^{*0}$) and
  $K^{*\pm}$ vector mesons in isobar (Ru+Ru and Zr+Zr) collisions. Results are compared
  with the same from Au+Au collisions. Right: Centrality dependence of
spin polarization ($\lambda_{\theta} \propto (3\rho_{00}-1)/(1-\rho_{00})$) of J/$\psi$ in Pb+Pb collisions
at 5.02 TeV~\cite{alice_jpsi_spin}.}
\label{fig-4}       
\end{figure}

The spin alignment of vector mesons can provide informations
complementary to hyperon spin polarization. The global spin alignment
is quantified by the diagonal element of the unity trace Hermitian spin density
matrix. Among diagonal elements, only the
$00^{\mathrm{th}}$ component (called $\rho_{00}$) can be measured from the angular
distribution ($\theta^{*}$) of the decay daughter of the vector meson with respect to
the polarization axis. The $\rho_{00}$ is expected to be $\frac{1}{3}$ in
absence of spin alignment. Any deviation of $\rho_{00}$ from $\frac{1}{3}$
indicates that there is a net spin alignment. 
In 2008, the measurements from 200 GeV Au+Au collisions by STAR resulted a null $\rho_{00}$ for
$K^{*0}$ and $\phi$ mesons with a large uncertainty~\cite{star_rho00_200}. Recent measurements by
ALICE in mid-central Pb+Pb collisions at LHC energies indicate that
the $K^{*0}$ $\rho_{00}$ significantly smaller than $\frac{1}{3}$
in the low $p_{\mathrm{T}}$ region, while the same for $\phi$ mesons
is consistent with $\frac{1}{3}$ within uncertainty~\cite{alice_kstar_phi_rho00}. At low $p_{\mathrm{T}}$, there is a clear
centrality dependence observed for both $K^{*0}$ and $\phi$. While the same
measurements in p+p collisions at LHC yielded null deviations in $\rho_{00}$.
The $\rho_{00}$ in Pb+Pb collisions is observed to be an order of
magnitude larger than the naive expectation  from $\Lambda$ polarization. 
The STAR collaboration reported of $\phi$ and $K^{*0}$ $\rho_{00}$
from the Beam Energy Scan~\cite{star_bes_rho00}. The BES data reveals a
surprising pattern. At mid-central collisions, while the $\phi$
$\rho_{00}$ is significantly larger than $\frac{1}{3}$, the $K^{*0}$ $\rho_{00}$
is consistent with $\frac{1}{3}$ within uncertainties. The large signal for $\phi$ mesons
can not be accommodated by conventional sources of
polarization, e.g. vorticity of the medium, electromagnetic field etc.
Polarization by a fluctuating vector meson strong force
field can explain the large deviation for
$\phi$ mesons in mid-central collisions~\cite{sheng_phi_field}. The high statistics isobar data from STAR allows a
precision measurement of $\rho_{00}$ for $K^{*0}$ and $K^{*\pm}$ and
presented in left panel in Fig.~\ref{fig-4}.  Due to a factor of five difference in magnetic moments
of $K^{*0}$ and $K^{*\pm}$ one naively expects $K^{*0}$ $\rho_{00}$ being larger
than $K^{*\pm}$ due to its coupling with $B$-field. But surprisingly
$K^{*\pm}$ $\rho_{00}$ is observed to be larger than $K^{*0}$ in
isobar collisions. The behind such an ordering is not understood yet
and require more inputs from theory.

Recently, the ALICE reported the first observation of $J/\Psi$ spin
polarization in di-muon channel at forward rapidity. The right panel in Fig.~\ref{fig-4}
presents the spin polarization parameter $\lambda_{\theta}$ ($\propto (3\rho_{00}-1)/(1-\rho_{00})$) as a
function of centrality in in Pb+Pb collisions at 5.02
TeV~\cite{alice_jpsi_spin}. On the contrary to $\rho_{00}$ of $\phi$
and $K^{*0}$ near mid-rapidity, the $J/\psi$ $\rho_{00}$ (translated
from $\lambda_{\theta}$) at forward rapidity is larger than $\frac{1}{3}$. This is
surprising and inputs from theory are needed for an explanation. 

\section{Summary}
\label{sec-2}

In summary, the spin polarization of hyperons across FAIR, RHIC and
LHC energies established the global nature of hyperon spin
polarization in heavy-ion collisions. More precise and differential
measurements will provide more insights and constrain various
models. Future $P_{\Lambda}$ measurements at the low energy will reveal whether or not
there is any vanishing polarization. A precise measurements of local
spin polarization ($P_{z}$) of hyperons are
carried out from RHIC to LHC. These results poses challenge to many
theories and provide informations on the complex structure of
vortices in heavy ion collisions. These measurements are generating
discussions of emerging spin phenomena (such as Shear Induced Polarization,
baryonic Spin Hall Effect) and help developing spin hydrodynamics,
spin kinetic theories etc. Measurements of spin alignments of vector mesons across RHIC and LHC
is very surprising and puzzling. The observation of opposite sign of $\rho_{00}$
deviations among different species requires more inputs from theory
for better understanding of the data. 

\section{Acknowledgement}
SS acknowledges support from the Strategic Priority Research Program of Chinese
Academy of Sciences (Grant XDB34000000).
%
%
%


\end{document}